\documentstyle[epsfig,12pt,preprint,tighten,aps]{revtex}
\begin{document}

\title{ \rightline{May 1999}
\rightline{{\tt UM-P-99/11}}
\rightline{{\tt RCHEP-99/05}}
\ \\
Maximal $\nu_e \to \nu_s$ solution to
the solar neutrino problem: just-so, MSW or energy independent?}

\author{R. M. Crocker, R. Foot and R. R. Volkas}

\address{School of Physics\\
Research Centre for High Energy Physics\\
The University of Melbourne\\
Parkville 3052 Australia\\
r.crocker, foot, r.volkas@physics.unimelb.edu.au}

\maketitle

\begin{abstract}
We examine the maximal $\nu_e \to \nu_s$ solution to the solar
neutrino problem. This solution can be motivated by the exact
parity model and other theories. The $\nu_e$ survival probability 
exhibits one of three qualitatively different behaviours depending on the 
value of $\Delta m^2$, viz. approximately energy independent, just-so or 
MSW. By the last of these we mean an enhanced night-time event rate due to
regeneration in the Earth. We study all of these possibilities in
the context of the recent SuperKamiokande data.
\end{abstract}

\section{Introduction}

Five experiments have measured solar neutrino fluxes that 
are significantly deficient
relative to standard solar model 
expectations\cite{solar,superK}. Four out of 
the five experiments find overall fluxes that are 
roughly $50\%$ of the theoretical predictions. (The Chlorine
experiment sees a larger deficit.) Maximal mixing 
between the electron neutrino and a
sterile flavour has been proposed as 
the underlying explanation for these observations\cite{flv}.
This oscillation mode produces a $50\%$ reduction in 
the day-time solar neutrino flux
for a large range of the relevant $\Delta m^2$ parameter:
\begin{equation}
10^{-3} > \frac{\Delta m^2}{{\rm eV}^2} \stackrel{>}{\sim} 
\ \text{few}\ \times 10^{-10}.
\end{equation}
The upper bound arises from the lack of $\overline \nu_e$ disappearence 
in the CHOOZ
experiment\cite{chooz}\footnote{Note that this entire range for $\Delta m^2$ does not necessarily lead to any inconsistency with bounds imposed by big bang nucleosynthesis\cite{footbb}.}, while the lower bound is 
a rough estimate of the transition region between
the totally averaged oscillation regime and the ``just-so'' regime.

The very special feature of maximal mixing between the $\nu_e$ 
and a sterile flavour is well motivated by the 
Exact Parity Model (also known as the mirror matter
model)\cite{flv,flv2}. 
In this theory, the sterile flavour maximally mixing with the $\nu_e$
is identified with the mirror electron neutrino. 
The characteristic maximal mixing feature occurs because 
of the underlying exact parity symmetry between the ordinary 
and mirror sectors. The potentially maximal mixing 
observed for atmospheric muon neutrinos
is beautifully in accord with this framework \cite{fvy}, 
which sees the atmospheric neutrino problem resolved 
through `$\nu_\mu \to$ mirror partner' oscillations.\footnote{The
mirror partners of the three ordinary neutrino flavours 
are distinct, effectively sterile light neutrino flavours.} 
{\it The Exact Parity Model therefore provides a
unified understanding of the solar and atmospheric 
neutrino problems: each is due to maximal oscillations of the 
relevant ordinary neutrino into its mirror partner.}
Note that the mirror neutrino scenario is phenomenologically 
similar to the pseudo-Dirac scenario\cite{pd}.

The chlorine result is not quantitatively 
consistent with this view of the origin
of the solar neutrino anomaly, it being about $30\%$ too low. 
We await with interest some new experiments that
have the capacity to double-check this result.

The purpose of this paper is to make a more detailed analysis of the maximal
$\nu_e \to \nu_s$ solution to the solar neutrino problem. We do so because of
two recent developments: (i) the clarification from Guth et al.\cite{guth}
that a day-night
asymmetry generically exists even for maximal mixing
(due to Earth regeneration which affects the night-time events)
and (ii) the observation by SuperKamiokande of an interesting feature 
in the recoil electron energy spectrum
for $E > 13$ MeV\cite{superK}. We will calculate the 
day-night asymmetry and the recoil electron spectrum
as a function of $\Delta m^2$ in the range $10^{-3}$ eV$^2$ to 
$10^{-11}$ eV$^2$. We will draw the important conclusion that 
the maximal $\nu_e \to \nu_s$ scenario has a larger number 
of characteristic and testable features than realised hitherto. 
We will summarise the ``smoking gun'' experimental 
signatures for this scenario in the concluding section.

\section{Day-night asymmetry}

Guth et al.\cite{guth}\ have recently provided a very 
lucid account of the physics of the day-night
effect for maximally mixed solar neutrinos. This is 
important for the maximal $\nu_e \to \nu_s$ scenario 
for two reasons. First, high statistics experiments such 
as SuperKamiokande have an on-going programme to
measure the solar neutrino day-night asymmetry. 
It had been previously and erroneously thought that 
maximally mixed $\nu_e$'s would {\it not} give rise to 
a day-night asymmetry.  We will calculate this asymmetry for 
the total, energy-integrated flux relevant for
SuperKamiokande as a function of $\Delta m^2$. We will 
see that the present data already rule out a range 
of $\Delta m^2$. If a nonzero asymmetry
were to be experimentally established in the future, 
then this would {\it not} falsify the maximal 
$\nu_e \to \nu_s$ scenario, contrary to previous belief.
Rather, such an observation would help to pin down the actual 
value of $\Delta m^2$. 

The second consequence of doing the physics correctly 
is the revelation that the night-time flux reduction is 
generically {\it not} an energy-independent factor
of a half as previously advertised. This is simply because 
the Earth regeneration effect \cite{bou} is energy-dependent. 
Data samples which do not separate day-time and night-time observations
would thus be expected to show a (weak) energy 
dependent suppression if the maximal
$\nu_e \to \nu_s$ scenario is correct.

Since the paper by Guth et al.\ provides a complete 
account of how the day-night asymmetry
is calculated, we will not repeat all of this material here. 
Suffice it to say that we 
checked our numerical procedure by recalculating some 
of the results given by Guth et al.\ for 
maximal $\nu_e \to \nu_{\mu}\ (\text{or}\ \nu_\tau)$ oscillations. 
We found agreement. The new computation
we present here is very similar, but 
involves changing the effective potential for 
neutrino oscillations in the Earth to the one relevant 
for $\nu_e \to \nu_s$ oscillations. It is given by
\begin{equation}
V = \sqrt{2} G_F (N_e - {N_n \over 2}) \simeq 
{G_F \over \sqrt{2}}{\rho \over m_N}(3Y_e - 1),
\end{equation}
where $G_F$ is the Fermi constant, $N_e$ ($N_n$) is 
the terrestrial electron (neutron) number 
density, $m_N$ is the nucleon mass, $\rho$ is the 
terrestrial mass density, and $Y_e$ is the number 
density of electrons per nucleon. We used the terrestrial 
density profile given in Ref.\cite{earth}.
The core of the computation is the determination of 
the recoil electron flux $g(\alpha, T)$ at
apparent recoil energy $T$ and 
zenith angle $\alpha$ (using the notation of Guth et al.). 
It is given by
\begin{equation}
g(\alpha, T) = \int_0^{\infty}dE_\nu \Phi(E_\nu) 
\int_0^{T'_{\text{max}}} dT' R(T,T')
{d\sigma \over dT'}(T', E_\nu, \alpha)
\end{equation}
where $\Phi(E_\nu)$ is the Boron neutrino spectrum \cite{bla}, $R(T, T')$ 
is the energy resolution function
given by \cite{res}
\begin{equation}
R(T,T') = {1 \over \Delta_{T'} \sqrt{2\pi}} 
\exp \left[- {(T' - T)^2 \over 2 \Delta^2_{T'}} 
\right],
\label{resolutionfunction}
\end{equation}
and the effective cross-section is given by \cite{bks}
\begin{equation}
{d\sigma \over dT'}(T', E_\nu, \alpha) = 
P(E_\nu,\alpha) {d\sigma_{\nu_e} \over dT'}(T',E_\nu).
\end{equation}
The function $P$ is the electron neutrino survival 
probability incorporating matter effects
in the Earth, and $d\sigma_{\nu_e}/dT'$ is the $\nu_e - e$ 
scattering cross-section. 
The resolution width is given by\cite{delta}
$\Delta_{T'}/\text{MeV} \simeq 0.47 \sqrt{T'/\text{MeV}}$.

Our result is presented in Fig.1, where the energy integrated 
day-night asymmetry $A_{n-d}$ is plotted as
a function of $\Delta m^2$ for maximal $\nu_e \to \nu_s$ 
oscillations. The asymmetry is defined by
\begin{equation}
A_{n-d} \equiv {  \int_{6.5\ \text{MeV}}^{\infty} dT 
\left[ N(T) - D(T) \right] \over
\int_{6.5\ \text{MeV}}^{\infty} dT \left[ N(T) + D(T) \right]    },
\end{equation}
where $6.5$ MeV is the energy threshold relevant for 
the day-night asymmetry measurement of 
SuperKamiokande, $D(T)$ is the day-time recoil
electron flux and $N(T)$ is the night-time flux. The day-night
asymmetry is computed from $g(\alpha, T)$ via the procedure 
described in Appendix B of Guth et al.
An average is performed over all zenith angles and seasons of the year.
Observe that the day-night asymmetry is positive and peaks with 
a value of about $20\%$ when $\Delta m^2 \simeq 10^{-6}$ eV$^2$. The 
present SuperKamiokande result\cite{superK},
\begin{equation}
A_{n-d} = + 0.026 \pm 0.021\ (\text{stat. + sys.}),
\end{equation}
yields a $2\sigma$ upper bound of roughly $A_{n-d} < 0.068$ which is 
plotted as a horizontal 
line in Fig.1. We see that the range
\begin{equation}
2 \times 10^{-7} \stackrel{<}{\sim} {\Delta m^2 \over \text{eV}^2} 
\stackrel{<}{\sim} 8 \times
10^{-6}
\end{equation}
is disfavoured at the $2\sigma$ level. The regions 
immediately to the side of the disfavoured
region will obviously be probed as more data are 
gathered. The asymmetry falls to the $1\%$ 
level at about $\Delta m^2 = 3 \times 10^{-8}$ eV$^2$ 
and $5 \times 10^{-5}$ eV$^2$.

If a positive nonzero value for $A_{n-d}$ were to be measured, 
then an ambiguity would
remain in the determination of $\Delta m^2$: values on either 
side of the presently
disfavoured region can produce the same $A_{n-d}$. 
This ambiguity could in principle be resolved from a determination of the
energy dependence of the night-time rate.
Figures 2a and 2b depict the energy dependence of the flux reduction for two
representative values of $\Delta m^2$ on either side of the disfavoured region. 
Figure 2a takes $\Delta m^2 \simeq 10^{-7}$ eV$^2$, 
with the solid (dotted) line showing the ratio of night-time (day-time)
flux per unit energy to the no-oscillation 
expectation. The dot-dashed curve is the
average. Note that the day-time rate is rigorously an 
energy-independent factor of two less
than the no-oscillation rate, while the night-time rate 
is $5-8\%$ higher than the day-time rate and is weakly 
energy-dependent. Figure 2b considers the same quantities for $\Delta m^2 =
10^{-5}$ eV$^2$. Note that the slopes of the night-time 
fluxes have opposite signs on
opposite sides of the disfavoured region. The energy dependence becomes
unobservably small for $\Delta m^2$ values away from the interval around
$10^{-6}$ eV$^2$.

\section{Recoil electron spectrum}

An interesting situation exists at present with regard to the recoil electron
energy spectrum measured by SuperKamiokande\cite{superK}. 
If the $hep$ neutrino flux predictions from standard
solar models are taken at face value, then 
SuperKamiokande has evidence for a distortion in
the boron neutrino induced recoil energy spectrum 
for energies greater than about 13 MeV.
We will call this the ``spectral anomaly''.
More specifically, the spectral anomaly is an excess of 
events relative to what would be expected
on the basis of an energy-independent boron neutrino 
flux reduction of about $50\%$. The observed
distortion also disfavours the popular small and 
large mixing angle MSW solutions
to the solar neutrino problem.

One can readily identify three possible interpretations of the spectral anomaly: (i) Standard 
solar models grossly underestimate the $hep$ 
neutrino flux. (ii) There is an as yet unidentified 
systematic error in the energy resolution function used 
by SuperKamiokande, and/or in their energy 
calibration. (iii) It is a statistical fluctuation.
(iv) New physics is the cause.
We will not consider (i) in this paper, and instead focus on (iv). Before doing so, we briefly
discuss (ii) as a cautionary note.

Figure 3 illustrates the effect of varying the 
resolution width $\Delta_{T'}$. We fit the data
by minimising the $\chi^2$ function
\begin{equation}
\chi^2 = \sum_{i=1}^{18} \left[ { N^{\text{exp}}_i - 
0.5 f N^{\text{th}}_i \over \sigma(N^{\text{exp}}_i) }
\right]^2 + \left[ {f - 1 \over \sigma(f)} \right]^2,
\label{chi2}
\end{equation}
where $N^{\text{exp}}_i$ is the measured recoil electron flux in 
energy bin $i$, $N^{\text{th}}_i$ is the
theoretical no-oscillation expectation for same, $0.5$ 
represents an energy independent $50\%$
suppression due (for instance) to averaged maximal 
$\nu_e \to \nu_s$ oscillations and $f$ is a boron
neutrino flux normalisation parameter to take account 
of the theoretical uncertainty 
$\sigma(f) \simeq 0.19$ in this quantity \cite{bla}. 
We include the two low energy bins, $5.5-6.0\ \text{MeV}$ and
$6.0-6.5\ \text{MeV}$, as well as the 16 other energy bins
used by SuperKamiokande \cite{superK}.
Seasonal and daily 
averages are taken. These data are fitted by varying $f$ 
and the quantity $\Delta$ defined through
\begin{equation}
\Delta_{T'} = (\Delta\ \text{MeV})\sqrt{{T' \over \text{MeV}}}.
\end{equation}
We find the minimum at $f = 0.90$ and $\Delta \simeq 0.51$, with 
$\chi^2_{\text{min}} \simeq 20$ for $19-2=17$ degrees of freedom, 
which is a good fit. The spectral anomaly becomes an artifact
of the finite resolution of the detector. SuperKamiokande 
quote a central value for $\Delta$ of about $0.47$\cite{res}. 
We can see from Fig.3 that a $5-10\%$ systematic shift 
upward of $\Delta$ would remove the anomaly. It is interesting 
to compare this result with Fig.1 from Ref.\cite{sm} which shows
the effect of an unidentified systematic error in the 
energy calibration of the detector. Systematic errors in
either or both of the energy resolution and 
energy calibration can in principle explain the
spectral anomaly. Having sounded this cautionary note, 
we now proceed on the basis that SuperKamiokande have in 
fact correctly determined their energy resolution and calibration
capabilities, and that new physics is behind the spectral anomaly.
 
It is easy to understand how the spectral anomaly can be 
explained in a very amusing way through maximal $\nu_e \to \nu_s$ 
oscillations. It could be that $\Delta m^2$ is such
that the transition energy between the averaged oscillation 
regime and the just-so regime is about 13 MeV\cite{bfl}! The spectral 
anomaly might represent the onset of just-so behaviour.
This would of course be a great piece of luck, but we 
probably should not disregard the possibility just out of 
world weary pessimism. To determine the $\Delta m^2$ 
value required to achieve this effect, the
appropriate $\chi^2$ function is minimised by 
varying $\Delta m^2$ and $f$.
The appropriate
$\chi^2$ is similar to Eq.(\ref{chi2}), 
except that $0.5 N^{\text{th}}_i$ is replaced by
a properly computed convolution integral. 
The $\Delta m^2$ dependence of $\chi^2$ is
shown in Fig.4. There are deep local minima in 
the $4 \times 10^{-10}$ eV$^2$ to $10^{-9}$ eV$^2$ 
region, which then trail off into a flat $\chi^2$ as the averaged
oscillation regime is entered. The minimum $\chi^2$ 
value is about 17 at $\Delta m^2 \simeq 6 
\times 10^{-10}$ eV$^2$ (and $f = 0.92$), which for
17 degrees of freedom represents a excellent 
fit. This $\Delta m^2$ can in principle be 
further probed through the anomalous seasonal effect. 
Preliminary calculations employing the $\Delta m^2$
which minimises $\chi^2$ show that an effect of  
about $6\%$ in magnitude can be obtained for the 
near-far asymmetry for certain of the 
higher energy bins of SuperKamiokande. We hope to 
return to an analysis of the seasonal effect in a
later paper.

\section{Conclusion}

Maximal $\nu_e \to \nu_s$ oscillations have been proposed as 
a solution to the solar neutrino
problem\cite{flv,pd}. This idea can be well-motivated by 
the mirror matter hypothesis\cite{flv}, or
by the pseudo-Dirac neutrino idea\cite{pd}. In this paper 
we have demonstrated the following points:
\begin{enumerate}
\item Across the allowed $\Delta m^2$ spectrum, maximal $\nu_e \to \nu_s$ oscillations lead to three qualitatively different behaviours for the $\nu_e$ survival probability; just-so, MSW or approximately energy independent. These occur in the ${\Delta m^2}/{\text{eV}^2}$ ranges
$10^{-3} > \text{energy independent} > 
\ \text{few} \times 10^{-5} >  \text{MSW} > \text{few} \times 10^{-8} >
\text{energy independent} > 10^{-9} > \text{just-so} > 
\ \text{few}\ \times 10^{-10}$. 
The day-night asymmetry observable provides for an experimental probe in the MSW regime. The recoil electron energy spectrum and 
the anomalous seasonal effect likewise provide a probe in the just-so regime.

\item Day-night asymmetry data rule out the range 
$2 \times 10^{-7} - 8 \times 10^{-6}$
eV$^2$ for $\Delta m^2$ at the $2\sigma$ level. 
A positive measurement of
the day-night asymmetry would pin down the 
preferred value to two possibilities on
either side of the disfavoured region, and the energy 
dependence of the night-time rate
could in principle resolve the ambiguity.
\item The SuperKamiokande spectral anomaly can 
be explained in this scenario if
$\Delta m^2$ takes any of the three or four 
values corresponding to the local
minima in Fig.4.
\item We have shown that 
increasing SuperKamiokande resolution width by $5-10\%$ would
also explain the spectral data.
\end{enumerate}

Finally, The $\Delta m^2$ region $10^{-3} - 10^{-5}$ eV$^2$ region 
can be probed through
the $\overline \nu_e$ disappearence experiment KAMLAND\cite{kam}
as well as through the atmospheric neutrino experiments\cite{fv}.
Note from the above that 
the region $\text{few}\ \times 10^{-8} - 10^{-9}$ eV$^2$ is without a 
smoking-gun signature. Of course, another
important test of the scenario for the entire $\Delta m^2$ 
range of interest will be
the neutral to charged current ratio at SNO\cite{sno}: they 
should obtain the no-oscillation value.
In addition, the BOREXINO \cite{ali} and iodine \cite{pal} experiments will 
double check the greater than $50\%$
suppression result distilled from the chlorine experiment.

\section{Acknowledgements}

RMC is supported by the Commonwealth of Australia and the 
University of Melbourne. RF
and RRV are supported by the Australian Research Council.

\section{Figure Captions}
\vskip 0.5cm
\noindent
Figure 1: Night-day asymmetry versus $\Delta m^2/\text{eV}^2$ (solid line)
for maximal $\nu_e \to \nu_s$ oscillations. Also shown is the night-day 
asymetry for maximal $\nu_e \to \nu_{\mu}$ oscillations for comparison
(dashed-dotted line). The horizontal dashed line is the $2\sigma$ upper limit.
\vskip 0.5cm
\noindent
Figure 2: Predicted recoil electron energy spectrum normalized
to the no oscillation expectation. Figure 2a (2b) is for 
$\Delta m^2/\text{eV}^2 = 10^{-7}$ ($10^{-5}$).
The solid (dotted) line corresponds to the ratio of night-time (day-time)
flux per unit energy to the no-oscillation 
expectation and the dot-dashed line is the average. 
\vskip 0.5cm
\noindent
Figure 3: The effect of varying the energy resolution, 
parameterized by $\Delta$ (see text).
The figure shows that the spectral anomaly can be explained
if $\Delta \simeq 0.50 \ \text{MeV}$ instead of the assumed 
SuperKamiokande value of $\Delta \simeq 0.47 \ \text{MeV}$.
\vskip 0.5cm
\noindent
Figure 4: Fit to the SuperKamiokande recoil electron
energy spectrum using maximal $\nu_e \to \nu_s$ oscillations.

\newpage
\epsfig{file=fig1.eps,width=15cm}
\newpage
\epsfig{file=fig2a.eps,width=15cm}
\newpage
\epsfig{file=fig2b.eps,width=15cm}
\newpage
\epsfig{file=fig3.eps,width=15cm}
\newpage
\epsfig{file=fig4.eps,width=15cm}

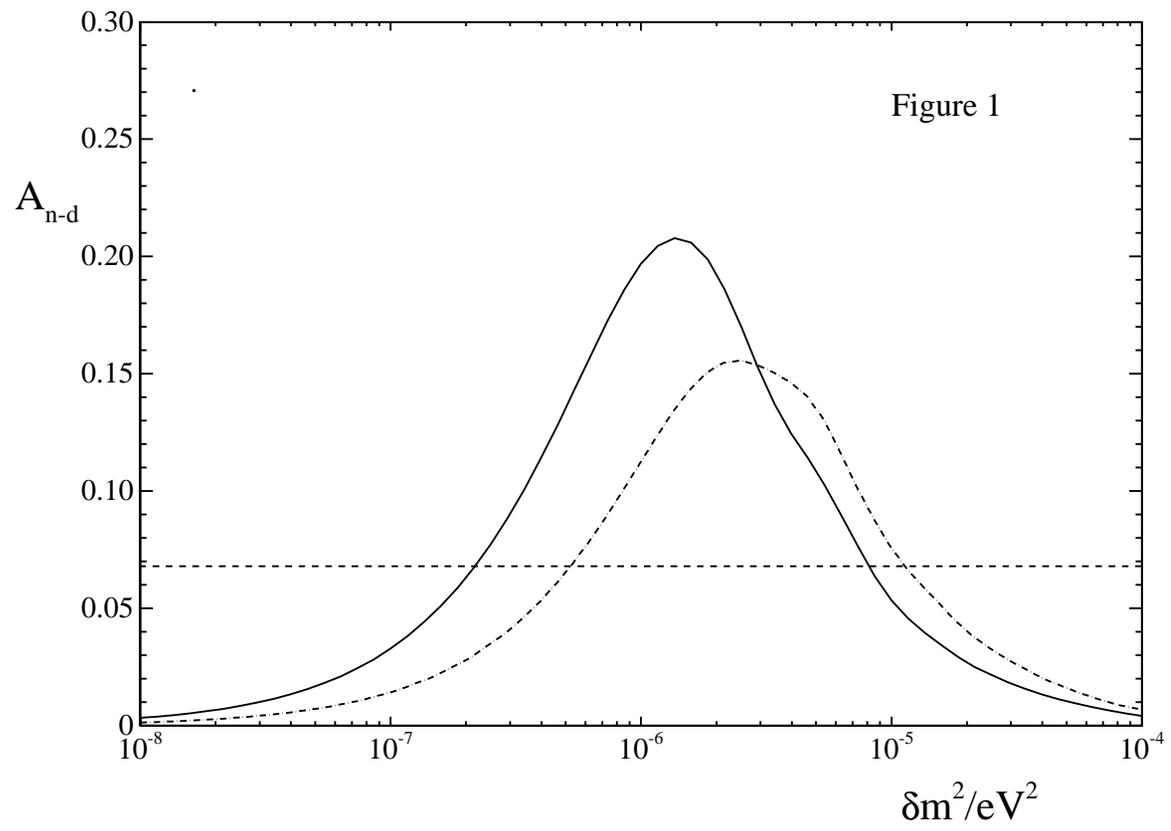
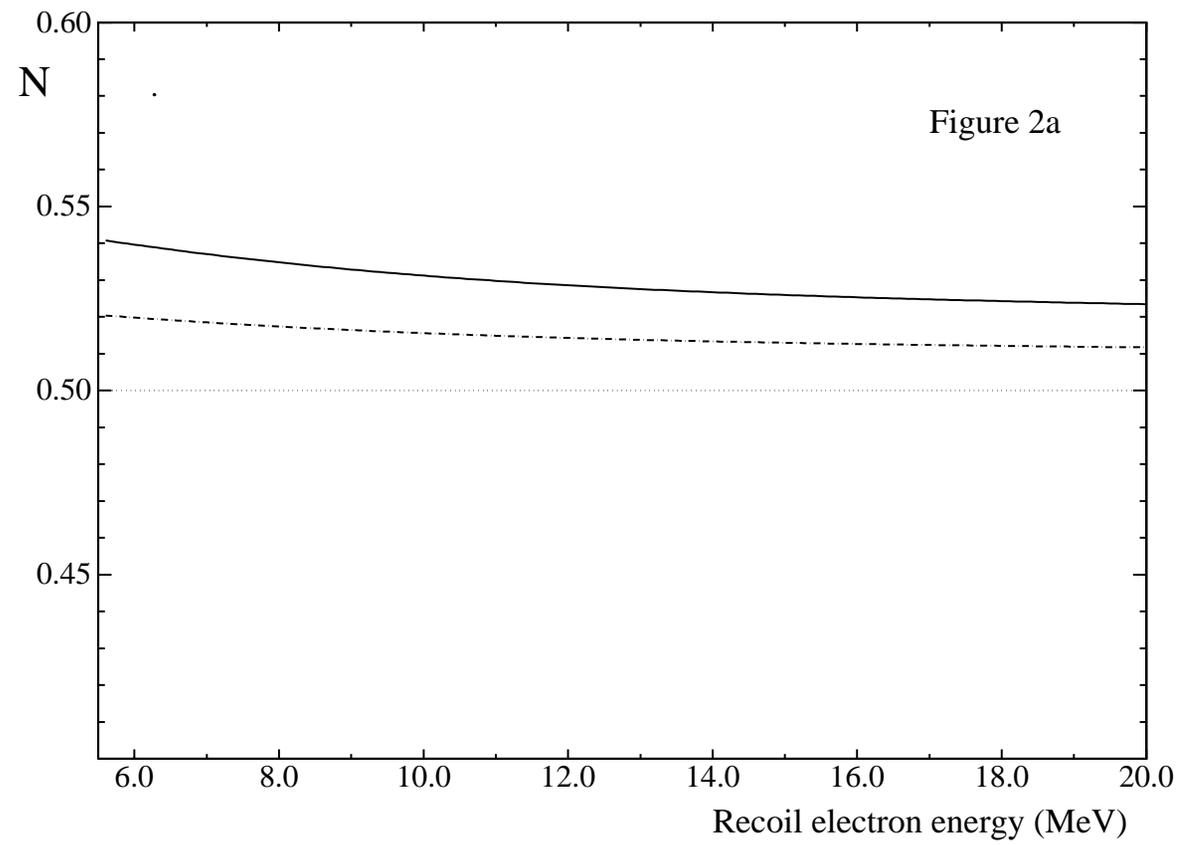
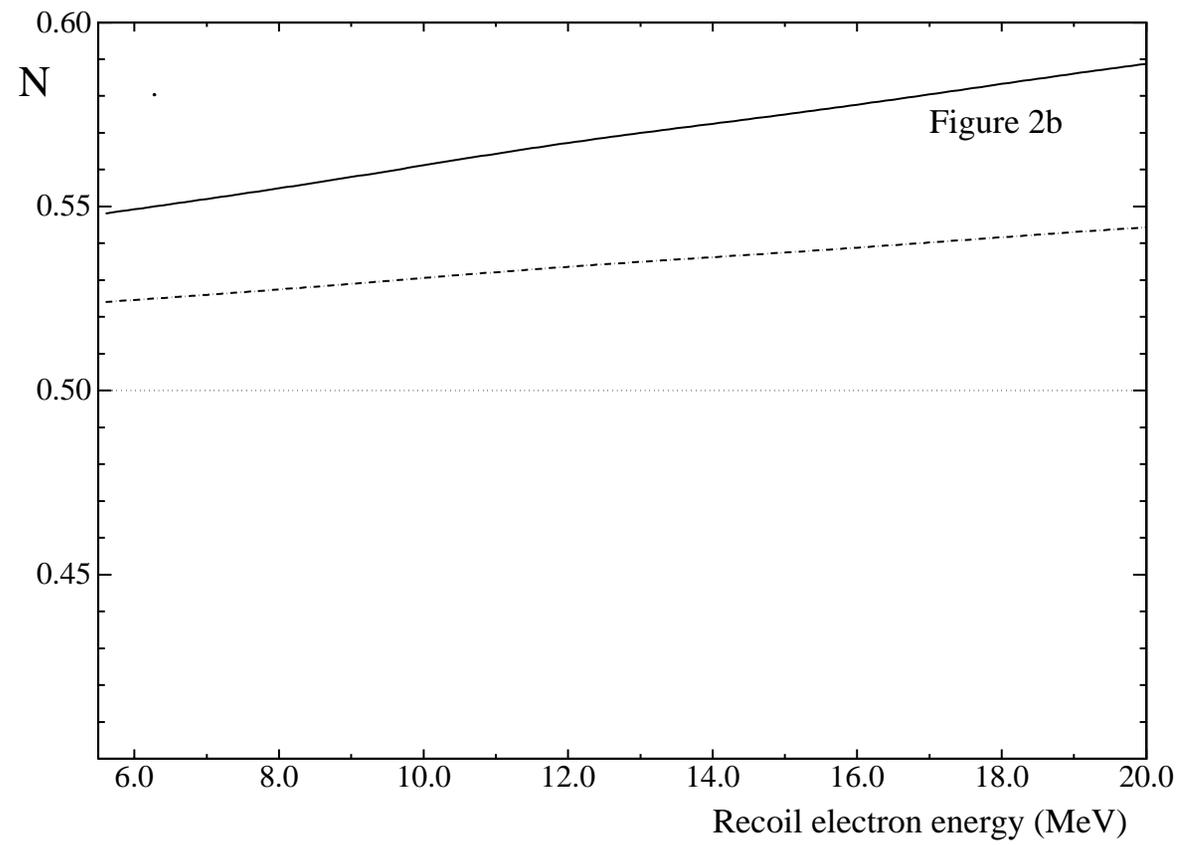
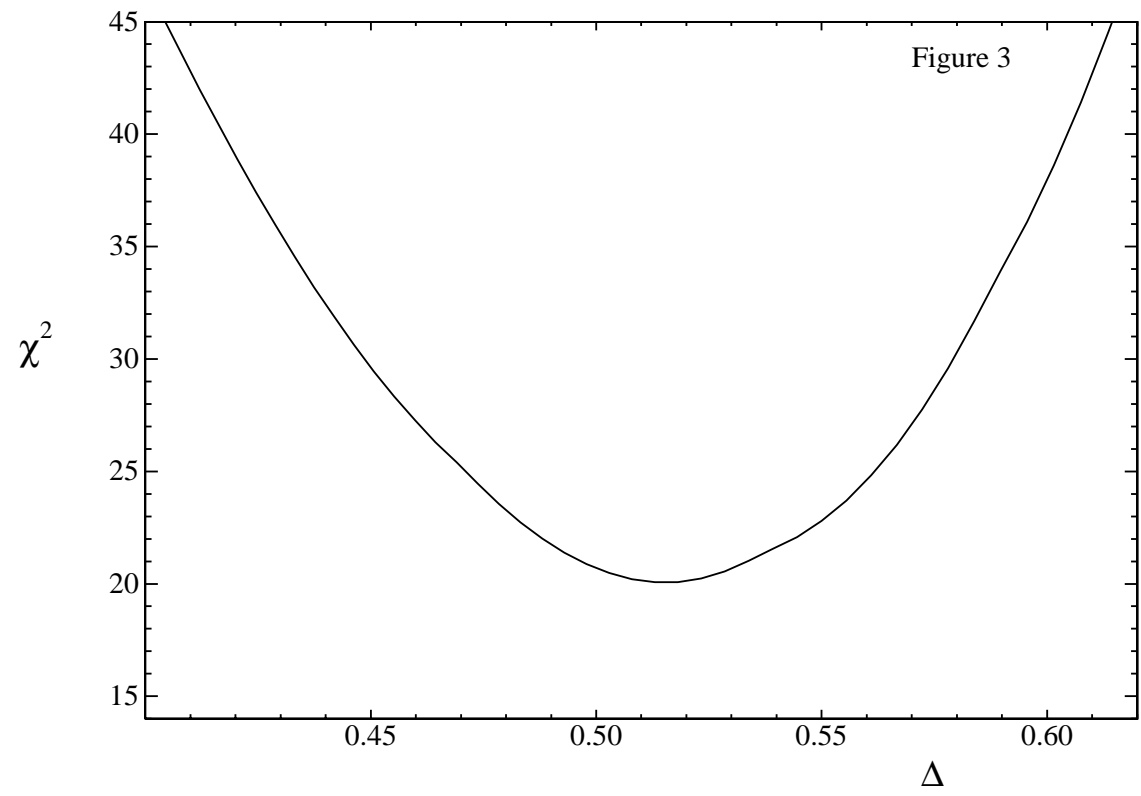
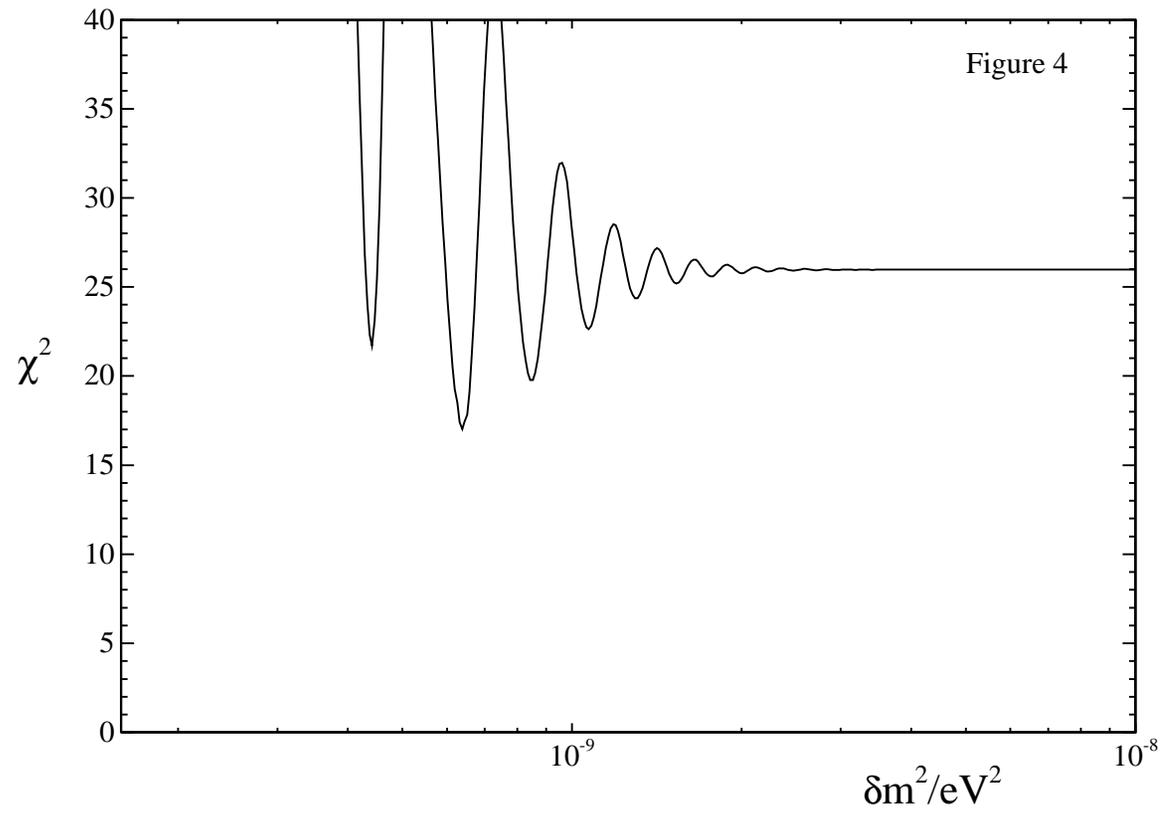
\end{document}